\documentclass[12pt]{article}
\usepackage[dvips]{graphicx}
\usepackage{amssymb}
\usepackage{amsmath}
\usepackage{here}

\newcommand{\bear}{\begin{array}}  
\newcommand {\eear}{\end{array}}
\newcommand{\bea}{\begin{eqnarray}}   
\newcommand{\eea}{\end{eqnarray}}
\newcommand{\beq}{\begin{equation}}   
\newcommand{\eeq}{\end{equation}}
\newcommand{\bef}{\begin{figure}}  \newcommand 
{\eef}{\end{figure}}
\newcommand{\bec}{\begin{center}}  \newcommand 
{\eec}{\end{center}}

\begin{document}

\begin{titlepage}

\begin{flushright}
ICRR-Report-587-2011-4\\
IPMU 11-0089 \\
UT-11-19
\end{flushright}

\vskip 2.0cm


\begin{center}

{\large \bf
Cosmological effects of decaying cosmic string loops with \\TeV scale width
}

\vskip 1.2cm

Masahiro Kawasaki$^{a,b}$,
Koichi Miyamoto$^a$
and
Kazunori Nakayama$^{c,d}$

\vskip 0.4cm

{\it $^a$Institute for Cosmic Ray Research,
University of Tokyo, Kashiwa 277-8582, Japan}\\
{\it $^b$Institute for the Physics and Mathematics of the Universe,
TODIAS,\\
University of Tokyo, Kashiwa 277-8568, Japan}\\
{ \it $^c$Department of Physics, University of Tokyo, Bunkyo-ku, Tokyo 113-0033, Japan}\\
{ \it $^d$ Theory Center, KEK, 1-1 Oho, Tsukuba, Ibaraki 305-0801, Japan}\\
 
 \date{\today}

\begin{abstract} 
In supersymmetric theories, cosmic strings produced in the early 
Universe often have a width of TeV scale,
while the tension is much larger.
In a scaling regime, an infinite cosmic string releases significant 
fraction of its energy in the form of string loops.
These thick string loops lose their energies efficiently by particle emissions, 
and hence it may have effects on cosmological observations.
We study cosmological implications of string loops with TeV scale width 
in detail and derive constraints on the tension of the string.
Implications on future gravitational wave detectors are also discussed.
\end{abstract}



\end{center}
\end{titlepage}

\section{Introduction}

Although the cosmological evolution scenario in the early Universe well 
before the big-bang nucleosynthesis (BBN) is not known yet, it is 
expected that the early Universe experienced some stages of spontaneous 
symmetry breaking (SSB) as the cosmic temperature decreases.
In association with the SSB, cosmic strings may be 
formed~\cite{Kibble:1976sj} depending on the topology of 
the vacuum manifold~\cite{Vilenkin}.
Besides, cosmic superstrings emerge at the end of the brane 
inflation~\cite{Sarangi:2002yt,Dvali:2003zj}.
Cosmic (super)strings, if detected, will provide a way to search for 
the early Universe and high-energy physics.
Their properties and cosmological consequences have been widely 
considered in many papers.

On the other hand, supersymmetry (SUSY)~\cite{Martin:1997ns} is one of 
the promising candidates of the physics beyond the standard model, since 
it provides a natural solution to the gauge hierarchy problem.
SUSY opens up a window toward the energy scale of the grand unified 
theory (GUT), and hence we can consistently discuss the physics of the 
early Universe in the framework of SUSY.
Formations of cosmic strings in the framework of SUSY GUT were 
discussed in the literature~\cite{Jeannerot:2003qv}.
Interestingly, a SSB which produces cosmic strings may be associated 
with SUSY breaking.
Suppose that there is a flat direction $\phi$ in the scalar field space,
which is ubiquitous in tree-level supersymmetric theories,
and that it has the soft SUSY-breaking negative mass-squared
$-m^2|\phi|^2$ around the origin.
Then, the vacua deviate far from the origin of the field space 
and the scalar fields have large nonzero vacuum expectation 
values (VEVs).
The width of the cosmic string is roughly determined by the curvature 
of the scalar potential at the origin and the tension is roughly 
given by the square of the VEV~\cite{Coleman}.
Since the energy scale of the soft SUSY breaking term, $m$, is of 
$\mathcal O$(TeV), the width of the cosmic string is estimated as 
$\sim (\rm{TeV})^{-1}$, unlike the ordinary GUT string whose width 
is expected to be about $(10^{16} \rm{GeV})^{-1}$.
The VEV, on the other hand, is determined by the balance between 
the negative soft mass term and higher dimensional operator, 
which is suppressed by the cutoff scale $M$, which can be as large as 
the Planck mass $M_{p}$.
In this case two mass-dimensional parameters which appear in the 
potential of the flat direction have extremely large hierarchy.
The VEV can be larger than $m$ by many orders of magnitude.
This may push up the tension of cosmic strings to 
a cosmologically relevant value.
   
A similar situation has been considered in the context of thermal 
inflation~\cite{Yamamoto:1985rd,Lyth:1995hj,Lyth:1995ka,Asaka:1999xd}.
Thermal inflation, which is the short-lasting inflationary period 
introduced as a solution to the cosmological moduli 
problem~\cite{Coughlan:1983ci,Banks:1993en}, requires the scalar 
field (called {\it flaton}) with light mass and large VEV 
in order to get a sufficient amount of e-folds to dilute away the moduli.
It is often assumed that the flaton have negative soft mass term 
at the origin, and is stabilized at a large field value
due to non-renormalizable terms.
If it has a U(1) symmetry that is broken due to the flaton VEV,
cosmic strings are produced at the end of thermal inflation with 
properties just described above.

The cosmic strings with TeV-scale width and much higher scale tension 
have characteristic cosmological effects beyond which ordinary strings 
have, such as gravitational waves (GWs), anisotropy of cosmic microwave 
background (CMB) and so on.
In particular, their extreme thickness enhances the particle production 
at a cusp on a loop.
A cusp is a spiky and extremely Lorentz-boosted region on a cosmic 
string loop which appears several times per one period of 
the oscillation of the loop.
When a cusp appears, two parts on the both sides of the cusp overlaps 
each other and this induces nonperturbative production of the scalar 
particles~\cite{Brandenberger:1986vj,BlancoPillado:1998bv}.
This effect clearly becomes more efficient as the width of the string 
get thicker.
The produced scalar particles decay to lighter particles in 
the standard model sector.
If this happens during the BBN epoch, it can disastrously modify 
the light element abundances.
Besides, the decay products may contain stable particles, which may be 
a candidate of dark matter (DM), 
and may provide too much matter abundance to be consistent with 
current observations~\cite{Jeannerot:1999yn,Cui:2008bd}.
These cosmological effects from the cusp annihilation crucially 
depends on the typical loop size created per Hubble time,
parameterized by $\alpha$, which is highly uncertain up to now.
These observations motivate us to study cosmological effects of 
TeV-width strings in detail, and obtain constraints on 
the parameter space of the loop size $\alpha$ and the tension $\mu$ 
through GWs, DM and BBN.
Cosmic strings associated with a flat direction were considered 
in several works~\cite{Freese:1995vp,Barreiro:1996dx,Penin:1996si,
Perkins:1998re,Cui:2007js} and especially Ref.~\cite{Cui:2007js} 
partly referred to the cosmological constraint from such cosmic strings.
In this paper, we further investigate detailed 
constraints to $\alpha$ and $G\mu$ of cosmic strings with TeV scale
width from various cosmological effects and also argue detectability 
of GWs emitted from these strings at future experiments.

This paper is organized as follows. 
In Sec.~\ref{sec:model} a simple SUSY model for the spontaneous $U(1)$
breaking is presented and show that cosmic strings with TeV scale 
width naturally appears.
In Sec.~\ref{sec:decay} evolution of string loops are briefly shown 
taking account of the GW emission and particle emission from cusps.
In Sec.~\ref{sec:effect} we discuss how the particle emission and GWs 
from string loops affect cosmological observations.
In Sec.~\ref{sec:constraint} we show constraints on the string tension 
and typical loop size, and also model parameters.
We conclude in Sec.~\ref{sec:conc}.

\section{Cosmic string models with TeV scale width}  
\label{sec:model}

Let us consider superfields $\phi_+$ and $\phi_-$, which have 
the charge $+1$ and $-1$ under the additional $U(1)$ gauge symmetry, respectively.
One of the candidates of an additional $U(1)$ is the $U(1)_{\rm B-L}$ symmetry, 
but we do not specify a concrete setup.
We also assume that there is R-symmetry and both $\phi_+$ and $\phi_-$ have R-charge $+1/n$ ($n$ is a positive integer).
These assumptions prohibit any renormalizable terms including 
$\phi_+,\phi_-$ in the superpotential and the allowed 
non-renormalizable term is 
\beq
	W=\frac{(\phi_+\phi_-)^n}{nM^{2n-3}},    \label{WNR}
\eeq
where $M$ is the cut-off scale of the low-energy effective theory.
The scalar potential is given by
\begin{equation}
	V = V_F + V_D+V_{\rm soft},
\end{equation}
where
\begin{equation}
	V_F = \frac{1}{M^{4n-6}}|\phi_+|^{2n-2} |\phi_-|^{2n-2} 
	(|\phi_+|^2+|\phi_-|^2),
\end{equation}
\begin{equation}
	V_D = \frac{g^2}{2}( |\phi_+|^2-|\phi_-|^2 )^2,
\end{equation}
\begin{equation}
	V_{\rm soft} = - m_+^2|\phi_+|^2 - m_-^2 |\phi_-|^2 
	- \left( A\frac{(\phi_+\phi_-)^n}{nM^{2n-3}} +{\rm h.c.}\right).
\end{equation}
where $g$ is the coupling of the additional $U(1)$ and soft masses 
$m_+$ and $m_-$ are expected to be of order of TeV.
$A$ is also of order of TeV and we have implicitly redefined scalar 
fields such that $A$ is real and positive.
Here, we use same letters for superfields as their scalar components. 
We assume that the soft mass terms for $\phi_+$ and $\phi_-$ is negative.
The D-term potential forces scalar fields to the D-flat direction, $|\phi_+|=|\phi_-|$.
The F-term and soft SUSY breaking terms produce the global minimum 
in this direction at\footnote{
Strictly speaking, the true global minimum shifts from the D-flat 
direction if $m_+ \ne m_-$.
However, the soft term is subdominant to the D-term 
around (\ref{minimum}), then the shift is small.
}
\bea
	 |\phi_+|  = |\phi_-|
	   &= &\left[ \frac{M^{2n-3}}{4n-2}
	 \left( A + \sqrt{A^2 + (4n-2)(m^2_+ + m^2_-)}\right)
	 \right]^{1/(2n-2)}, \\[0.6em]
	 \arg \phi_+ &= & -\arg \phi_-. \label{minimum}
\eea
Parametrically,
\beq
	 v\simeq (mM^{2n-3})^{1/2(n-1)}, \label{mu_of_M}
\eeq
where $v$ is the scale of the VEVs of the scalar fields and $m$ is 
the scale of the coefficients of soft SUSY breaking terms, that is, 
$m\sim m_+ \sim m_- \sim A$.
Since the vacuum manifold is $S^1$, this potential leads to 
the emergence of cosmic strings if the $U(1)$ is restored at first 
and then spontaneously broken after that.
The width of cosmic string $w$ is roughly given by the inverse of 
the curvature of the scalar potential at the origin as
\beq
	 w \simeq m^{-1} \sim (\rm{TeV})^{-1}.
\eeq
The tension $\mu$ is roughly given by the square of the VEV 
of the scalar field, namely
\beq
	\mu \simeq v^2 \sim (mM^{2n-3})^{1/(n-1)}.
\eeq
The feature $\sqrt{\mu} \gg m$ does not depend on the detailed model construction.
It is rather a generic feature as long as the soft SUSY breaking mass triggers the SSB.
If we take  $M\sim M_p$ and $m\sim \rm{TeV}$, 
$\mu \sim (10^{10}\rm{GeV})^2$ for $n=2$ and  
$\mu \sim (10^{14}\rm{GeV})^2$ for $n=3$.
These correspond to $G\mu \sim 10^{-18}$ and $G\mu \sim 10^{-10}$, respectively.
Although these values are much smaller than the current upper limit
$(G\mu \lesssim 10^{-7})$, cosmic strings with such a tension still 
deserve cosmological interest, as discussed below. 

The potential naturally causes a short-period of secondary inflation, 
called thermal inflation~\cite{Yamamoto:1985rd,Lyth:1995hj,
Lyth:1995ka,Asaka:1999xd}.
During inflation, $\phi_+$ and $\phi_-$ get masses of order of 
the Hubble parameter through supergravity effect,
\beq
	V_H=H_{\rm inf}^2(c_+|\phi_+|^2+c_-|\phi_-|^2),
\eeq
where $c_+$ and $c_-$ are $\mathcal{O}(1)$ constants and assumed 
to be positive here, and $H_{\rm inf}$ is the Hubble parameter 
during inflation.
If $H_{\rm inf}  > m$, which is satisfied in most inflation models, 
$\phi_+$ and $\phi_-$ are trapped at the origin during inflation.
After inflation, the energy stored in the inflaton is converted 
to the thermal bath through the reheating process.
At the origin, the $U(1)$ is restored and hence the $U(1)$ gauge 
filed is massless.
Then there appears finite temperature corrections to the scalar 
potential as
\beq
	V_T=c_T T^2(|\phi_+|^2+|\phi_-|^2).
\eeq
Here $c_T$ is a $\mathcal{O}(1)$ positive coefficient.
This stabilizes $\phi_+$ and $\phi_-$ at the origin for $T \gtrsim m$.
When the temperature falls below $T_{\rm be}\sim \sqrt{mv}$, 
the potential energy of the scalar fields ($\sim m^2v^2$) exceeds 
the energy of the thermal bath and thermal inflation starts.
During this period the preexisting radiation and matter is diluted exponentially 
and its temperature goes down.
The unwanted relics such as the gravitino and moduli are diluted away,
and hence it is appealing for solving the cosmological moduli problem~\cite{Coughlan:1983ci,Banks:1993en}.
When it falls down to $T_{\rm end}\sim m$, the finite temperature 
correction becomes negligible and the true potential minimum appears.
Then the scalar field rolls down toward the potential minimum and thermal inflation ends.
After thermal inflation, cosmic strings are formed since the $U(1)$ 
symmetry is spontaneously broken by the VEVs of $\phi_{\pm}$.

The flaton decays after thermal inflation, and the radiation dominated universe restarts.
The reheating temperature after thermal inflation must exceed a few 
MeV for the successful BBN~\cite{Kawasaki:1999na}.
The decay rate of the flaton is estimated as
\beq
	\Gamma=\gamma \frac{m^3}{v^2},
\eeq
where $\gamma$ is a numerical constant and the reheating temperature is given by
\bea
   T_{\rm RH} & = & \left(\frac{90}{\pi^2 g_*}\right)^{1/4}
   (M_p\Gamma)^{1/2} \nonumber \\
   & \simeq & 100 {\rm MeV} \left(\frac{g_*}{10}\right)^{-1/4} 
   \left(\frac{\gamma}{0.1}\right)^{1/2} 
   \left(\frac{v}{10^{14}\rm{GeV}}\right)^{-1}
   \left(\frac{m}{1\rm{TeV}}\right)^{3/2},
   \label{T_RH}
\eea
where $g_*$ is the number of relativistic degrees of freedom at $T=T_{\rm RH}$.
For example, the flaton couples to the Higgses as~\cite{Martin:1996kn}
\begin{equation}
	W = \lambda \frac{\phi_+\phi_-}{M_P}H_u H_d,
\end{equation}
yielding a right magnitude of the higgsino mass of $\mu_H = \lambda v^2 / M_P$.
In this case, the flaton decays into Higgs boson pair and we obtain $\gamma \sim (\mu_H / m)^4$
and hence $\gamma$ takes rather wide range of values depending on $\mu_H$.
For $m=1~\rm{TeV}$ and $\gamma=0.1$, requiring 
$T_{\rm RH}\gtrsim \mathcal{O}({\rm MeV})$ leads to
$v\lesssim 10^{16} {\rm GeV}$ and  we need $v\lesssim 10^{14}$~GeV
for $m=100~{\rm GeV}$ and $\gamma=0.1$.
This condition sets another upper bound for the tension of cosmic strings.

\section{Decay of cosmic string loops}     
\label{sec:decay}

Once formed, the distribution of cosmic strings in the universe 
obeys the so-called scaling law.
In the scaling regime, a few infinite cosmic strings exist 
per Hubble horizon.
In order for the strings to fall into the scaling regime, cosmic 
strings must collide and reconnect with each other and cast their energies 
into the string loops.
Cosmic string loops lose their energies by several processes,
which may leave characteristic signatures on cosmological observations.
We consider two mechanisms in which cosmic string loops lose 
their energies; GW emission and particle emission.

\begin{description}
\item[({\rm i})] Gravitational wave emission

GW emission from cosmic string loops~\cite{Damour:2000wa} 
is usually thought of as the main process of the energy losses of loops.
GW bursts are emitted mainly from cusps and kinks (discontinuous 
inflections on loops) and cusps are more efficient sources compared with kinks.
The energy emission rate of GWs of frequency $\omega$ from a cusp obeys a power law, 
$d\dot{E}/d\omega\propto \omega^{-4/3}$, and the total energy loss 
rate is found by summing up the emission rates of all modes as
\beq
	\dot{E}_{\rm GW}\simeq \Gamma G\mu^2, \label{P_GW}
\eeq
where $\Gamma\simeq 50$ is a numerical constant.

\item[({\rm ii})] Particle emission

When a cusp is formed on a loop, two string branches on each side of 
the cusp overlap.
It is expected that scalar particles are created due to non-perturbative effects.
Since it is difficult to estimate precisely the non-perturbative 
effect, we here assume that the conversion from the energy of 
the condensate into particles is efficient enough to approximate 
the total energy of the created particles to be that stored in the overlap region.
Then the energy loss rate of a loop through particle emission 
is estimated as~\cite{BlancoPillado:1998bv} (see also Appendix),
\beq
	\dot{E}_{\rm PE}\simeq p\mu\left(\frac{w}{l}\right)^{1/2}, 
	\label{P_PE}
\eeq
where $p$ is the number of time when cusps appear on a loop 
in one period of its oscillation, $w$ is its width and $l$ is 
its circumference.
Note that particle emission becomes more important for thicker 
or smaller loops.
This is simply because the overlap region near a cusp becomes 
relatively larger for such loops.
The exponents of $\mu$ in (\ref{P_GW}) and (\ref{P_PE}) imply 
that for smaller $G\mu$ particle emission becomes more important 
compared with GW emission.

\end{description}

\noindent
The loop size at which the efficiencies of these effects are equal 
is given by
\beq
l_= = w\left(\frac{p}{\Gamma G\mu}\right)^2.
\eeq
For $l<l_=$, a loop loses its energy mainly by particle emissions.

Cosmic string loops shrink by loosing their energy and finally disappear.
The loop length evolves according to
\beq
\mu \frac{dl}{dt} = -\Gamma G\mu^2 - \mu p \sqrt{\frac{w}{l}}.
\eeq
This can be solved as
\beq
   t-t_i=\frac{l_=}{\Gamma G\mu}\left[
   \left(\frac{l_i-l}{l_=}\right) 
   - 2\left(\sqrt{\frac{l_i}{l_=}}-\sqrt{\frac{l}{l_=}}\right) 
   + 2\ln \left(\frac{1+\sqrt{l_i/l_=}}{1+\sqrt{l/l_=}}\right)\right], 
   \label{l_of_t}
\eeq
where $t_i$ is the time at the birth of the loop and $l_i$ is 
the initial loop size.
Here we make an assumption that the initial loop size created 
at the cosmic time $t_i$ is given by $l_i = \alpha t_i$, with $\alpha$ 
being a constant smaller than unity.
The magnitude of $\alpha$ remains unknown.
Its suggested value ranges from $0.1$~\cite{Vanchurin:2005pa,Olum:2006ix} 
to some powers of $G\mu$~\cite{Siemens:2002dj,Polchinski:2006ee}.
Intermediate or mixture results are shown in Refs.~\cite{Martins:2005es,
Ringeval:2005kr,Dubath:2007mf,Vanchurin:2010me}.
Considering these uncertainties, we will treat $\alpha$ as 
a free parameter.

Let us consider some limiting cases.
If $l\gg l_=$, GW emission dominates the energy loss and then 
we have $\dot{l}=-\Gamma G\mu$.
Therefore, loops become shorter at the constant rate : 
\beq
	l - l_i = -\Gamma G\mu(t-t_i)  {\rm ~~~for~~~}  l\gg l_=.
\eeq
This is also directly obtained from (\ref{l_of_t}).
If $\alpha =l_{i}/t_{i}\ll \Gamma G\mu$, a loop disappears soon after it is born 
and if $\alpha \gg \Gamma G\mu$, it survives much more than 
one Hubble time.
On the other hand, if $l\ll l_=$, we obtain 
$\dot{l}=- p \sqrt{\frac{w}{l}}$.
Therefore, we obtain
\begin{equation}
	\left( \frac{l}{l_=} \right)^{3/2} 
	-\left( \frac{l_i}{l_=} \right)^{3/2} 
	=-\frac{3\Gamma G\mu}{2l_=}(t-t_i)   {\rm ~~~for~~~}  l\ll l_=.
\end{equation}
The fractional change of the loop length in one Hubble time at the 
birth is roughly given by
\beq
	\frac{\dot{l}t}{l} \simeq p\frac{t}{l}\sqrt{\frac{w}{l}}
\eeq
and this becomes unity when
\beq
	\frac{l}{t} \simeq \left(\frac{p^2 w}{t}\right)^{1/3}.
\eeq
Therefore, if $\alpha \ll (p^2 w/t_{i})^{1/3}$ a loop disappears within 
one Hubble time 
and if $\alpha \gg (wp^2 /t_{i})^{1/3}$ it takes longer than one Hubble time 
for a loop to disappear.

Particle emissions from cusps extract extra energy from loops, 
then the lifetime of a loop becomes shorter.
This alters the distribution of loops in the universe and as a result 
interesting cosmological signatures may be implied as shown below.

\section{Effects of cosmic string loops on the universe}  
\label{sec:effect}

\subsection{Abundance of cosmic string loops}

As explained, we parametrize the typical length of the string loop 
$l$ by a constant $\alpha$@such that the loop length is $\alpha t_i$ 
at the production time $t_i$.
The cosmic string network reaches the scaling regime, where there is 
$\mathcal{O}(1)$ infinite strings in each Hubble horizon.
In this regime, each infinite string abandons its large portion 
in the form of loops in each Hubble horizon per one Hubble time.
Therefore, the number density of loops at $t$ which are born between 
$t_i$ and $t_i+dt_i$ is given by
\beq
	\frac{dn(t, t_i)}{dt_i}dt_i \sim \alpha^{-1}t_i^{-4} 
	\left(\frac{a(t_i)}{a(t)}\right)^3dt_i, 
	\label{loopdensity}
\eeq 
at the cosmic time $t$.
Here, $(a(t_i)/a(t))^3$ represents the dilution due to the cosmic 
expansion.
The length of a loop at the time $t$ is found by solving (\ref{l_of_t}).

\subsection{Gravitational waves}   \label{sec:GW}

GW bursts from cusps on loops overlap each other and form 
the stochastic GW background.
The spectrum of GW background was calculated in 
Refs.~\cite{Damour:2000wa,Caldwell:1991jj,DePies:2007bm,Damour:2004kw,
Siemens:2006yp,Olmez:2010bi}.
In the calculation of the GW background, we should be careful that 
the GW background consists of only bursts which overlap other ones. 
When the contributions of many bursts to the background are summed up, 
we must omit bursts which come to the observer solely 
(which we dub as ``rare bursts'').
This was pointed out in Refs.~\cite{Damour:2000wa,Damour:2004kw,
Siemens:2006yp} and we follow the formalism of 
Ref.~\cite{Siemens:2006yp}.

The amplitude of the GW background without omitting rare bursts is given by the integral
\beq
    \Omega_{\rm GW}(f)=
    \frac{4\pi^2}{3H_0^2}f^3\int dz\int dl h^2(f,z,l)\frac{d^2R}{dzdl},
\eeq
where $f$ is the frequency of the GW, $z$ is the redshift at which 
the burst occurs, $l$ is the loop size, $h(f,z,l)$ is the present 
value of the strain of the GW emitted by the loop with size $l$ 
at redshift $z$, and $d^2R/dzdl\times dzdl$ is the burst rate 
which reach the observer emitted by loops with size $l\sim l+dl$ at redshift $z\sim z+dz$. 
$h(f,z,l)$ is given by~\cite{Damour:2000wa}
\beq
   h(f,z,l)\simeq C \frac{G\mu l}{((1+z)fl)^{1/3}}\frac{1}{fr(z)}, 
   \label{strain}
\eeq
where $C\simeq 2.68$ and $r(z)=a_0\int^{t_0}_{t(z)} a(t)^{-1}dt$ is 
the proper distance.   
$d^2R/dzdl$ is derived as follows.
The rate of bursts emitted at redshift $z\sim z+dz$ by loops 
which are born at $t_i\sim t_i+dt_i$ is~\cite{Damour:2000wa}
\beq
   \frac{d^2R}{dzdt_i}dzdt_i 
   = \frac{1}{4}\theta_m^2 \frac{2p}{(1+z)l(t(z),t_i)}
   \frac{dn}{dt_i}\frac{dV}{dz}\Theta(l(t(z),t_i))dzdt_i,   
\eeq
where $\theta_m=((1+z)fl)^{-1/3}$ is the extent of the burst, 
$dV/dz\times dz$ is the proper spatial volume between the redshifts 
$z$ and $z+dz$, $l(t,t_i)$ is given as a solution of (\ref{l_of_t}) 
and $\Theta$ is the step function.
We get $d^2R/dzdl$ by converting variables from $(z,t_i)$ to $(z,l)$ 
using (\ref{l_of_t})
\beq
   \frac{d^2R}{dzdl} = \frac{d^2R}{dzdt_i}
   \times \left[ 
   \left(1+\sqrt{\frac{l_=}{l}}\right)
   \left(\Gamma G\mu+\alpha\frac{1}{1+\sqrt{l_=/l_i}}\right)
   \right]^{-1}. 
   \label{dR_dzdl}
\eeq
Rare bursts are omitted from the sum in the following way.
Since stronger GWs are rarer, it is plausible to set the cut-off 
strain $h_*$ above which GWs are thought of as rare bursts. 
Then, we set $h_*$ from
\beq
R_{>h_*}=\int^{\infty}_{h_*}dh\int^{z_*}_0 dz \frac{d^2R}{dhdz} =f, \label{def_of_hstar}
\eeq
where $z_*$ is the redshift at which cosmic strings appear.
Eq.~(\ref{def_of_hstar}) means that GWs of larger strain than 
$h_*$ are observed less than once in a period of GWs themselves 
$f^{-1}$.
GWs of smaller strain than $h_*$ are observed many times 
during the period $f^{-1}$ and overlap each other.
Therefore, the stochastic GW background spectrum is estimated as
\beq
   \Omega_{\rm GW}(f)=\frac{4\pi^2}{3H_0^2}f^3
   \int^{h_*}_0 dhh^2\int^{\infty}_0 dz \frac{d^2R}{dhdz}. 
   \label{Omega_GW}
\eeq
 

\begin{figure}[tbp]
\begin{center}
\includegraphics[width=110mm]
{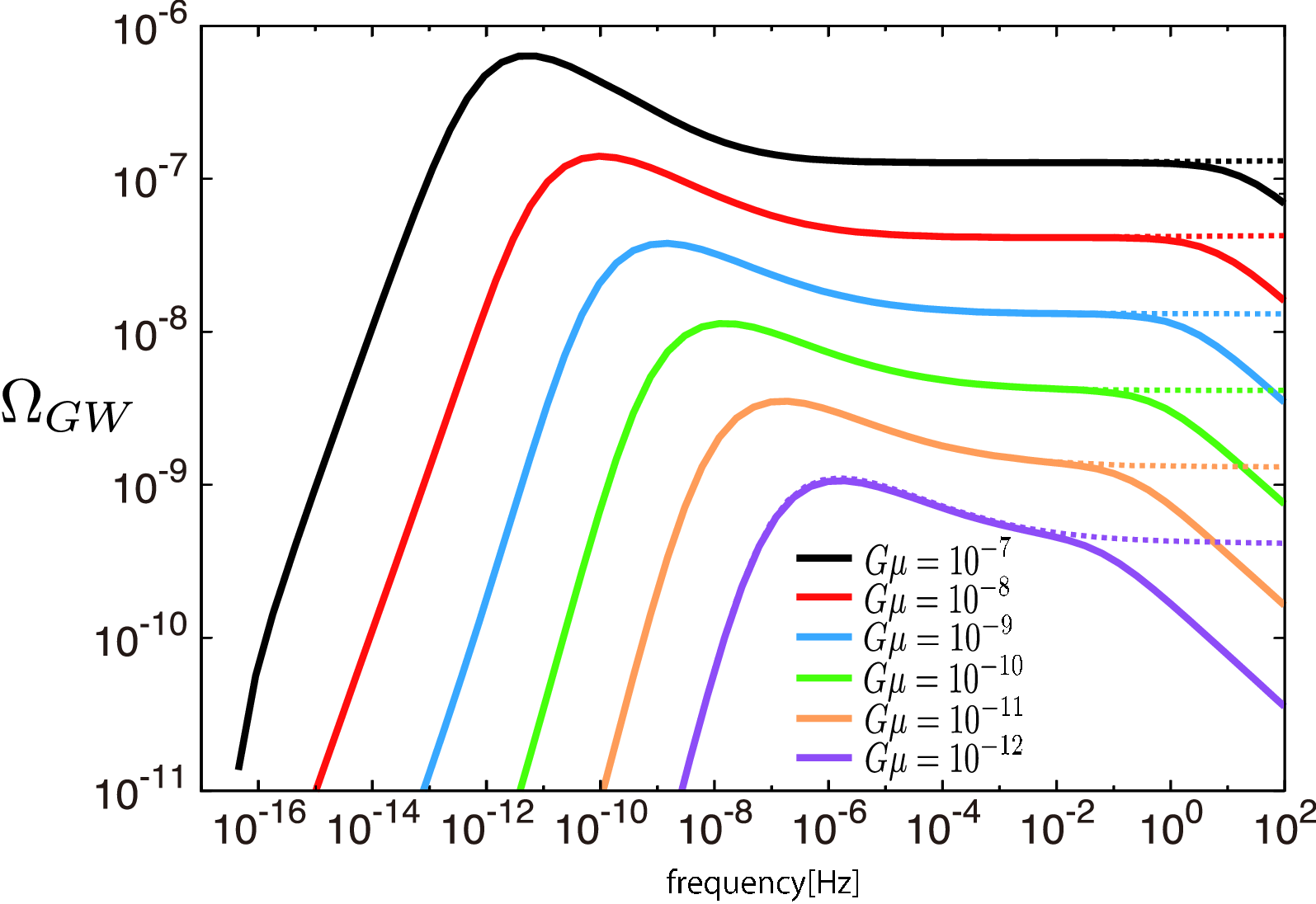}
\caption{The spectrum of the GW background, $\Omega_{\rm GW}(f)$, 
for various values of $G\mu$ with $\alpha=0.1$.
Solid lines take into account the effect of particle emission 
and dotted ones do not.}
\label{fig:OmegagwGmu}
\end{center}
\end{figure}



\begin{figure}[H]
\begin{center}
\includegraphics[width=110mm]
{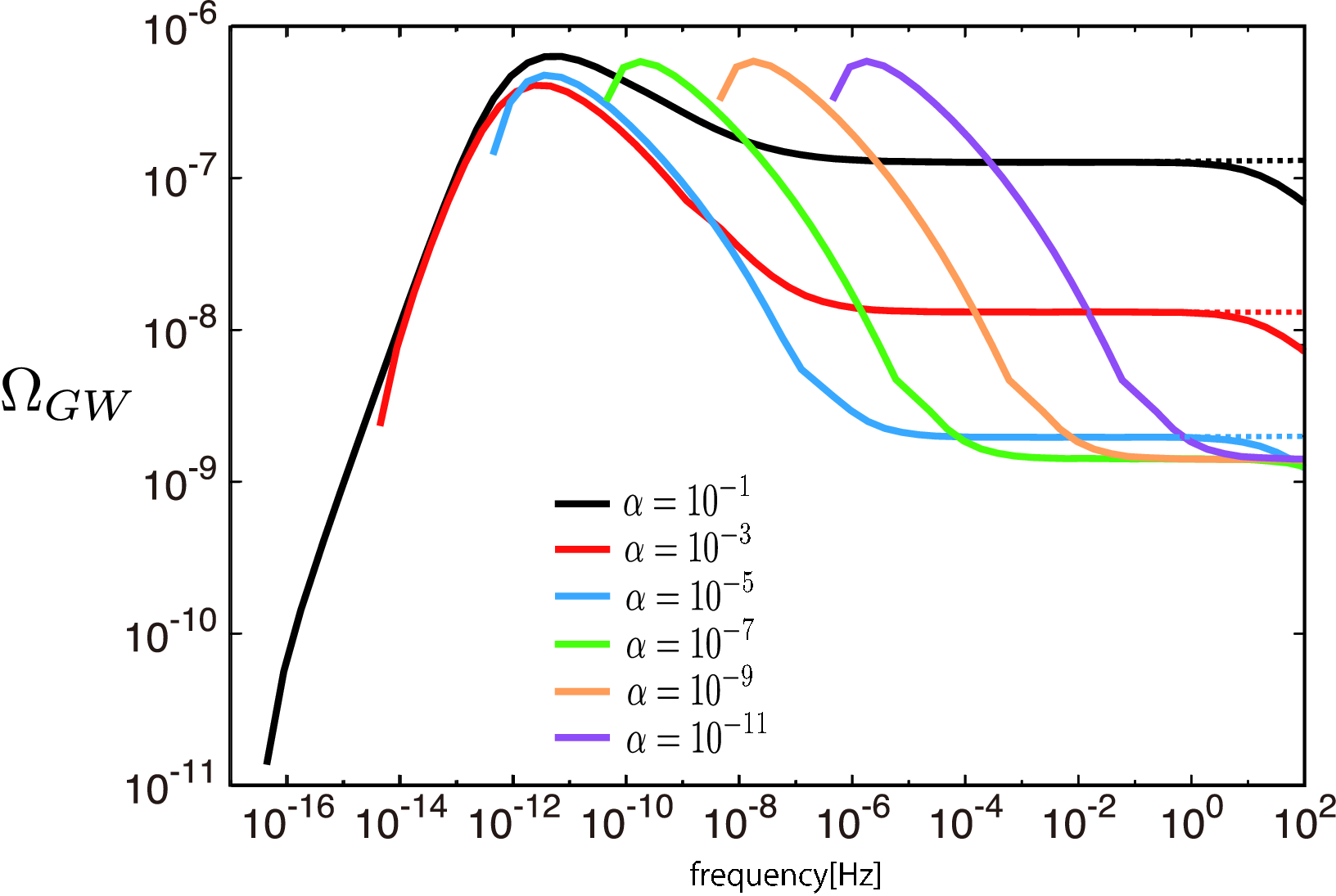}
\caption{The spectrum of the GW background, $\Omega_{\rm GW}(f)$, 
for various values of $\alpha$ with $G\mu=10^{-7}$. 
Solid lines take into account the effect of particle emission 
and dotted ones do not.}
\label{fig:Omegagwalpha}
\end{center}
\end{figure}


We show the spectrum of the GW background for various values of $G\mu$ 
with $\alpha=0.1$ in Fig.~\ref{fig:OmegagwGmu}, and for various values of
$\alpha$ with $G\mu=10^{-7}$ in Fig.~\ref{fig:Omegagwalpha}. 
Other parameters are taken to be $w=(1{\rm TeV})^{-1}, p=1, \Gamma=50$.
The solid lines take into account the effect of particle emission. 
Dotted ones do not include it, which correspond to $w=0$.
The shapes of the solid lines are identical to those shown in previous 
papers~\cite{DePies:2007bm,Siemens:2006yp}.
GWs of higher frequencies are emitted earlier.
The plateau in the high frequency region corresponds to GWs emitted 
in the radiation-dominated era and the downward-sloping region 
corresponds to those from the matter-dominated era.
In the high frequency region, it is seen that $\Omega_{\rm GW}$ 
calculated including particle emission is suppressed compared with 
that neglecting it.
This is because such high frequency GWs are emitted at the time 
when the size of loops 
is so small that the energy of loops is converted into particles 
more efficiently.
The suppression is milder than shown in Ref.~\cite{Cui:2007js}, 
since in Ref.~\cite{Cui:2007js} it is assumed that when a loop becomes 
smaller than $l_=$ it instantly disappears converting its whole energy 
to particles.
On the other hand, we here trace the evolution of loop size even after 
particle emission dominates GW emission, and such a loop contributes 
to the GW background until it completely disappears.

We see in Fig.~\ref{fig:Omegagwalpha} that reducing $\alpha$ shifts 
the spectrum toward higher frequency direction and suppresses 
the amplitude.
The reason why the whole spectrum goes right is simply that
loops become small and unable to emit low frequency GWs.
The lower cut-off of the frequency of the GW background from loops 
is given by $\sim(\alpha t_0)^{-1}$ where $t_0$ is the present age 
of the universe.
GWs of frequencies $\omega\sim(\alpha t_0)^{-1}$ are emitted by loops 
born within one present Hubble time.
When $\alpha$ is extremely small, there are no GWs of low frequency 
emitted by loops, and the GW background in the low frequency region 
consists of GWs from kinks on infinite strings~\cite{Kawasaki:2010yi}. 
The amplitude in the plateau becomes larger as $\alpha$ gets larger 
by the following reason.
Loops emitted from infinite strings behave like matter, and their 
energy density decreases proportional to $a^{-3}$.
Therefore, the ratio of the energy of loops born 
in the radiation-dominated era to that of the radiation increases 
as time goes by until loops completely decay.
As a result, the energy fraction of GWs from loops become larger.

\subsection{Big-bang nucleosynthesis}  
\label{sec:BBN}

String loops are continuously produced from infinite strings during 
BBN epoch.
It is expected that scalar particles emitted from cusps on loops 
soon decay into lighter particles producing many hadrons, since 
the lifetime of the $\phi$ particle must be shorter than $1$~sec 
in order for successful reheating, as already explained.
Since they emit high energy particles, the standard BBN prediction 
on the light element abundances may significantly be affected.
Constraints on the energy injection from BBN were studied in 
Ref.~\cite{Kawasaki:2004qu,Jedamzik:2004er} and the most stringent 
bound comes from overproduction of primordial D or $^7$Li due to 
hadro-dissociation of $^4$He, which leads to 
$\Delta\rho_{\rm vis}/s \lesssim 10^{-14}{\rm GeV}$ 
at $T\sim 10~{\rm keV}$, where $\Delta\rho_{\rm vis}$ is the injected 
energy density per Hubble time and $s$ is the entropy density.

The energy injection from cusps in one Hubble time around 
$T\sim 10{\rm keV}$ is estimated as
\beq
   \Delta\rho_{\rm vis} \simeq 
   \int^{t_{\rm BBN}}_0 dt_i \frac{dn(t_{\rm BBN},t_i)}{dt_i}  
   \mu\sqrt{\frac{w}{l(t_{\rm BBN},t_i)}}  
   \Theta(l(t_{\rm BBN},t_i)) \times t_{\rm BBN},
\eeq
where $t_{\rm BBN}$ is the cosmic time at $T= 10~{\rm keV}$ and 
$dn(t_{\rm BBN},t_i)/dt_i$ is given by Eq.~(\ref{loopdensity}).
Thus BBN provides a constraint on the parameters of string loops.

\begin{figure}[H]
\begin{center}
\includegraphics[width=110mm]{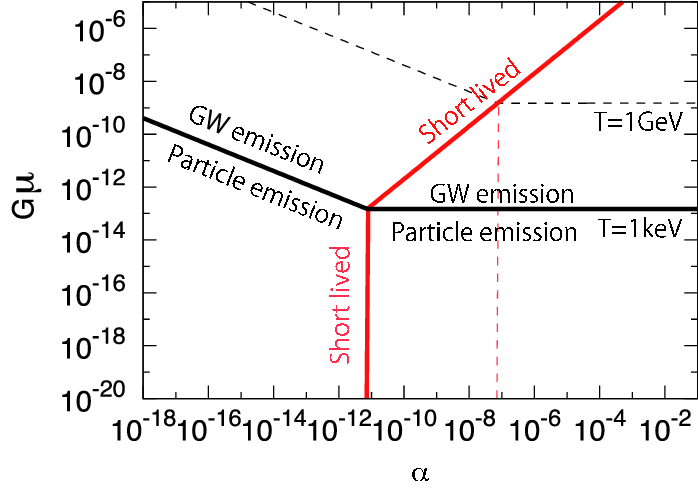}
\caption{
   Parameter regions where loops lose energies mainly through GW 
   emissions and particle emissions, and where loops are short-lived 
   and long-lived for $w=(1{\rm TeV})^{-1}$.
   Thick-solid (thin-dashed) lines correspond to loops created 
   at $T=1$keV (1GeV).
 }
\label{fig:cases}
\end{center}
\end{figure}

Let us focus on some limiting cases.
We can classify properties of string loops depending on whether 
GW emission dominates particle emission or not, and whether loops 
survive more than one Hubble time or not.
The loops disappear as soon as they are produced for 
$\alpha \ll \Gamma G\mu$ if GW emission dominates, and
for $\alpha \ll (p^2 w/t)^{1/3}$ if particle emission dominates, 
for the loops created at $t$.
On the other hand, the GW emission is dominant for 
$\Gamma G\mu > p(w/\alpha t)^{1/2}$ if loops soon disappear, 
and for $\Gamma G\mu > p(tp/w)^{1/3}$ if loops are long lived 
for loops disappearing at $t$.
These are summarized in Fig.~\ref{fig:cases}.

(1) Short-lived and GW emission dominated loops :
This corresponds to the upper-left region in Fig.~\ref{fig:cases}.
The energy density of the string loops is comparable to that of 
the infinite strings with a scaling regime,
\begin{equation}
	\rho_{\rm loop}(t) \simeq G\mu \rho_{r}(t).   \label{short-loop}
\end{equation}
The visible particle emission within a Hubble time per entropy 
density is estimated as
\begin{equation}
	\frac{\Delta \rho_{\rm vis}(t)}{s} 
	\simeq \frac{\rho_{\rm loop}(t)}{s} 
	    \frac{\dot E_{\rm PE}}{\dot E_{\rm GW}}
	\simeq  \frac{\rho_{r}(t)}{s} 
	    \left( \frac{p^2 w}{\Gamma^2 \alpha t} \right)^{1/2}.
\end{equation}

(2) Short-lived and particle emission dominated loops :
This corresponds to the lower-left region in Fig.~\ref{fig:cases}.
The energy density of the string loop is given by Eq.~(\ref{short-loop}).
Since the all of the loop energy goes into the visible sector, we obtain
\begin{equation}
	\frac{\Delta \rho_{\rm vis}(t)}{s} 
	\simeq G\mu \frac{\rho_{r}(t)}{s}.
\end{equation}

(3) Long-lived and GW emission dominated loops :
This corresponds to the upper-right region in Fig.~\ref{fig:cases}.
The main contribution to the loop energy density at the time $t$ 
comes from those disappearing at $t$, which were created at 
$t_i\sim \Gamma G\mu t /\alpha $.
Thus the energy density of the string loops exceeds that of 
the infinite strings with a scaling regime,
\begin{equation}
	\rho_{\rm loop}(t) 
	\simeq \rho_{\rm loop}(t_i)\left( \frac{a(t_i)}{a(t)} \right)^3 
	\simeq G\mu \rho_{r}(t)
	    \left( \frac{\alpha}{\Gamma G\mu} \right)^{1/2}. 
	\label{long-loop}
\end{equation}
Therefore, the visible particle emission from these loops is 
estimated as
\begin{equation}
	\frac{\Delta \rho_{\rm vis}(t)}{s} 
	\simeq \frac{\rho_{\rm loop}(t)}{s} 
	   \frac{\dot E_{\rm PE} t}{\mu l (t; t_i)}
	\simeq  \frac{\rho_{r}(t)}{s}
	   \frac{(\alpha w)^{1/2}}{\Gamma^2 G\mu t^{1/2}}.
\end{equation}

(4) Long-lived and particle emission dominated loops :
This corresponds to the lower-right region in Fig.~\ref{fig:cases}.
The dominant contribution to the loop energy density comes from 
those with the length of $l(t; t_i) \sim (p^2 t^2 w)^{1/3}$, 
which were born at $t_i \sim (p^2 t^2 w)^{1/3}/\alpha$.
The energy density of loops is given by
\begin{equation}
	\rho_{\rm loop}(t) 
	\simeq \rho_{\rm loop}(t_i)\left( \frac{a(t_i)}{a(t)} \right)^3 
	\simeq G\mu \rho_{r}(t)\left( \frac{\alpha^3 t}{p^2 w} \right)^{1/6}. 
\end{equation}
The visible particle emission is then estimated as
\begin{equation}
	\frac{\Delta \rho_{\rm vis}(t)}{s} 
	\simeq \frac{\rho_{\rm loop}(t)}{s} 
	    \frac{\dot E_{\rm PE} t}{\mu l (t; t_i)}
	\simeq  \frac{\rho_{\rm loop}(t)}{s}
	\simeq \frac{\rho_{r}(t)}{s}G\mu 
	    \left( \frac{\alpha^3 t}{p^2 w} \right)^{1/6}.
\end{equation}
From these expression, it is evident that the visible particle 
emission is efficient at earlier epoch.
The BBN constraint, however, is most stringent at $T\sim 10$~keV
and hence particle emission from loops are constrained from BBN 
dominantly at around $T\sim 10$~keV.


\subsection{Dark matter production} \label{sec:LSP}

It is expected that scalar particles emitted from cusps decay into 
standard model particles and SUSY particles with roughly equal branching 
ratio as long as the latter process is kinematically allowed.
At least two lightest supersymmetric particles (LSPs) are produced per 
$\phi$-decay into SUSY particle pair.
If the $R$-parity is conserved, the LSP is stable and contributes 
to the measured DM abundance.
This may explain DM or the condition that the density of this decay 
products must not exceed the current DM abundance sets constraints 
on the loop parameters.

Denoting the fraction of the energy that is converted into 
DM particles to the total energy emitted from cusps by
$\epsilon$, the LSP abundance is given by
\beq
   \Omega_{\rm LSP}=\frac{1}{\rho_{\rm cr}}\int^{t_0}_{t_*} dt_i 
       \int^{\bar{t}_{\rm co}(t_i)}_{t_{\rm fo}}dt
       \epsilon \frac{dn}{dt_i}(t,t_i) \mu \sqrt{\frac{w}{l(t,t_i)}} 
       \left(\frac{a(t)}{a(t_0)}\right)^3\Theta(t_{\rm co}-t_{\rm fo}),
    \label{eq:omega_lsp} 
\eeq
where $t_*$ is the cosmic time when cosmic strings appear, 
$t_{\rm fo}$ is the cosmic time when the LSP freezes out and 
$\rho_{\rm cr}$ is the critical density.
LSP freeze-out occurs when the temperature drops to 
$T_{\rm fo}\sim 10{\rm GeV}$.
The upper limit of the integral for $t$, the cosmic time when LSPs 
are emitted, is given by
\beq
   \bar{t}_{\rm co}(t_i)=
      \begin{cases}
          t_{\rm co}(t_i) \ & {\rm for}~~ \ t_{\rm co}<t_0 \\
          t_0               & {\rm for}~~ \ t_{\rm co}>t_0 ,
      \end{cases}
\eeq
where $t_{\rm co}(t_i)$ is defined by $l(t_{\rm co}(t_i),t_i)=0$.
In the integral the contribution from LSPs emitted before 
the freeze-out is omitted since they thermalize soon.
As described in Ref.~\cite{Cui:2008bd}, Eq.(\ref{eq:omega_lsp}) is 
rewritten as
\beq
   \Omega_{\rm LSP}=\frac{1}{\rho_{\rm cr}}\int^{t_0}_{t_*} dt_i 
      2\epsilon \frac{dn}{dt_i}(t_0,t_i)\mu l_=
      \left[
         \sqrt{\frac{l}{l_=}}-\ln \left(1+\sqrt{\frac{l}{l_=}}\right)
      \right]^{\bar{l}_{\rm fo}(t_i)}_{l_x(t_i)}
      \Theta(\bar{l}_{\rm fo}(t_i))\Theta(l_x(t_i)), 
   \label{OmegaDM}
\eeq
where
\beq
   \bar{l}_{\rm fo}(t_i)=
   \begin{cases}
      l(t_{\rm fo},t_i) & {\rm for}~~ \ t_i<t_{\rm fo} \\
      \alpha t_i        & {\rm for}~~ \ t_i>t_{\rm fo}
   \end{cases}
\eeq
and
\beq
   l_x(t_i)=
   \begin{cases}
      0          & {\rm for}~~ \ t_{\rm co}(t_i)<t_0 \\
      l(t_0,t_i) & {\rm for}~~ \ t_{\rm co}(t_i)>t_0.
   \end{cases}
\eeq
We expect that 
\beq
    \epsilon \sim \frac{m_{\rm LSP}}{m_\phi},
\eeq
where $m_{\rm LSP}$ is the mass of a LSP and $m_\phi$ is that of 
the scalar particle of which cosmic strings consist. 
Hereafter, we set $m_{\rm LSP}=100{\rm GeV}$, $m_{\phi}=1{\rm TeV}$ 
and $\epsilon=0.1$.
The condition that $\Omega_{\rm LSP}$ calculated by (\ref{OmegaDM}) 
does not exceed the current observed value 
$\Omega_{\rm DM}h^2 \simeq 0.1123$~\cite{Komatsu:2010fb} 
leads to the constraint on $\alpha$ and $G\mu$.

Analytic expressions for $\Omega_{\rm LSP}$
in some limiting cases are similar to those in the previous subsection,
except that we need to multiply an additional factor $\epsilon$ to them
and that they should be evaluated at the earliest possible epoch 
after DM decouples from thermal bath.
This is because the contribution on the DM abundance goes larger 
for loops created at earlier epochs~\cite{Cui:2008bd}.
Therefore, the requirement that $\Omega_{\rm LSP}$ should not 
exceed the current observed value limits the energy injection from 
cosmic strings around the LSP freeze-out, 
in contrast to the BBN constraint.

There is a subtlety which deserves to be commented about.
If the LSP freezes out in the radiation-dominated era, the redshift 
at the freeze-out is simply given by $1+z_{\rm fo} =T_{\rm fo}/T_0$, 
where $T_0$ is the today's temperature of the universe, 
and $t_{\rm fo}$ is simply obtained.
Analytic expressions for $\Omega_{\rm LSP}$ becomes similar to those 
in the previous subsection.
However, if the reheating temperature after the thermal inflation, 
which is given by (\ref{T_RH}), is lower than $T_{\rm fo}$, 
LSP freeze-out occurs while the oscillation of flaton dominates 
the universe.
As explained in \cite{earlyuniverse}, even during the oscillation 
of flaton, there is radiation whose temperature drops as 
$T\propto a^{-3/8}$.
Therefore, the redshift and the time at LSP freeze-out is given by 
\beq
   1+z_{\rm fo}
      = \left( \frac{T_{\rm fo}}{T_{\rm RH}}\right)^{8/3}
         (1+z_{\rm RH}), \ t_{\rm fo} 
      \simeq \left(\frac{1+z_{\rm RH}}{1+z_{\rm fo}}\right)^{3/2}
         t_{\rm RH},
\eeq
where $1+z_{\rm RH} = T_{\rm RH}/T_0$ and $t_{\rm RH}$ are 
the redshift and the time at the completion of reheating, respectively.
The dominant contribution on the DM abundance again comes from LSPs 
emitted around LSP freeze-out.
However, whether loops are short-lived or long-lived, the energy of 
LSPs at time $t$ is roughly given by 
\beq
    \Delta \rho_{\rm LSP}(t) 
       \sim G\mu \rho_{\rm tot}(t)
          \frac{\dot{E}_{\rm PE}}{\dot{E}_{\rm PE}+\dot{E}_{\rm GW}}, 
    \label{delrho_RH}
\eeq
where $\rho_{\rm tot}$ is the total energy density of the universe.
This is because the energy density of loops is not enhanced 
even if they are long-lived since both loops and oscillating flaton 
decrease their energy in proportion to $a^{-3}$.

\subsection{CMB anisotropy}

It is well known that the cosmic string network induces the CMB anisotropy.
However, observations such as WMAP have shown that cosmic strings 
cannot be the main source of the CMB anisotropy and hence the string 
contribution to the cosmic density fluctuation is constrained.
According to the recent calculation, this bound reads 
$G\mu \lesssim 5\times 10^{-7}$~\cite{Battye:2010xz}.
This is independent of the parameter $\alpha$.
Hereafter, we concentrate on the case where $G\mu$ is smaller than 
$10^{-7}$.

\section{Cosmological constraints}  \label{sec:constraint}

Now let us derive the constraints on the tension $(G\mu)$, and 
loop size ($\alpha$), and also find the parameter region where future 
experiments are sensitive.
We take account of constraints from BBN and LSP production 
as described in the previous section.
We include future or on-going GW detectors such as
ultimate DECIGO~\cite{Seto:2001qf}, BBO, BBO-correlated~\cite{BBO}, 
LISA~\cite{LISA} and advanced LIGO~\cite{LIGO} as the limits and 
sensitivities on GW emission.
We also use the current pulsar timing 
limit~\cite{Jenet:2006sv,vanHaasteren:2011ni} and future sensitivities 
from SKA~\cite{SKA} on the GW background from cosmic strings.
The sensitivity curves of future GW experiments are found 
in Ref.~\cite{Smith:2005mm}.

\begin{figure}[h]
\begin{center}
\includegraphics[width=120mm]{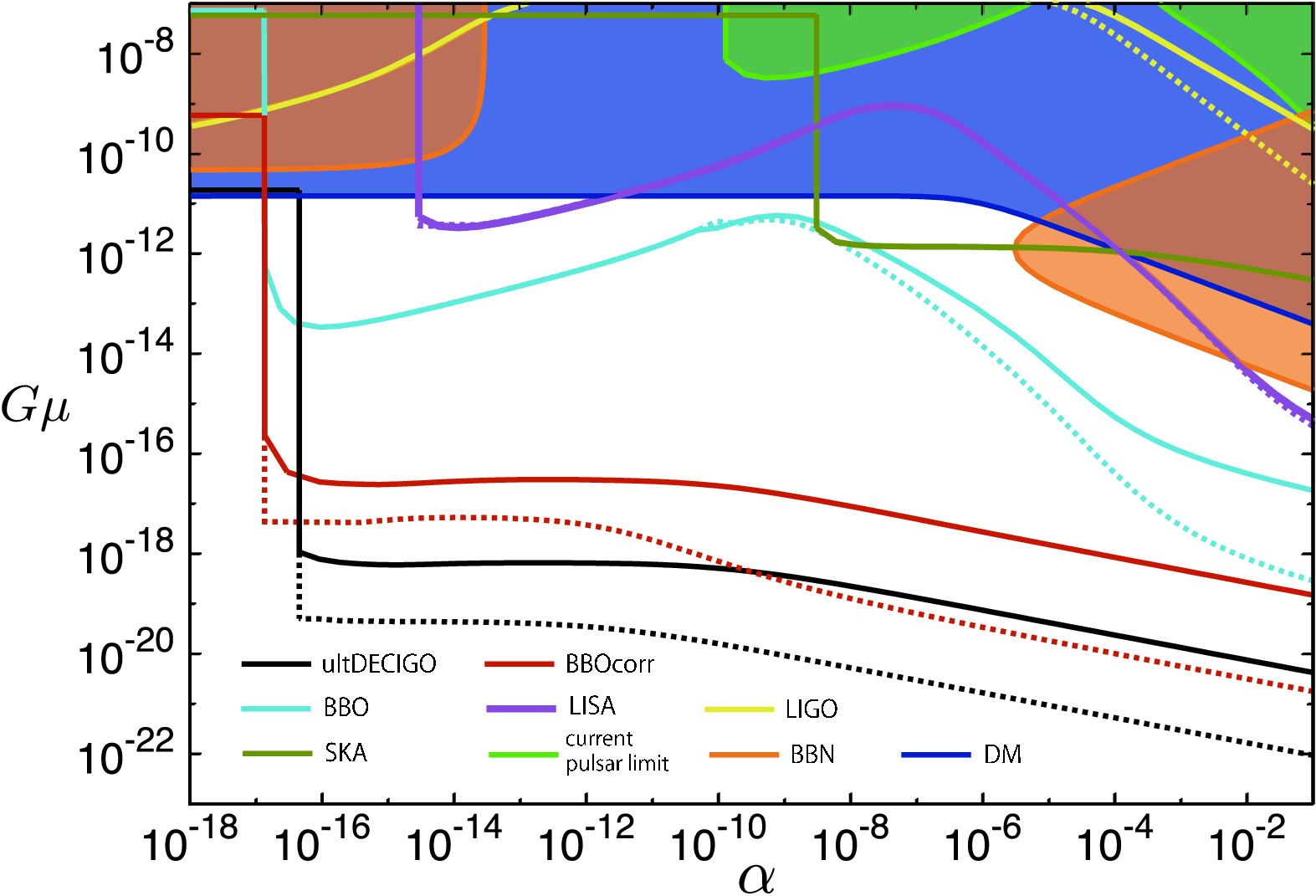}
\caption{
   Parameter regions excluded by cosmological constraints 
   and probed by future experiments on $\alpha$-$G\mu$ plane.
   Colored regions are excluded by current observational limits 
   or cosmological constraints.
   The area above each line can be probed by corresponding future 
   experiments.
   Solid lines take into account the effect of particle emission 
   into calculation of the GW background, while dotted lines do not. 
}
\label{fig:alphaGmuall}
\end{center}
\end{figure}

The result is shown in Fig.~\ref{fig:alphaGmuall}.
Solid lines take into account the effect of particle emission 
in the calculation of the GW background, while dotted lines do not.
The solid and dotted lines corresponding to sensitivities of LISA 
and SKA and the current limit from the pulsar timing overlap each other, 
so dotted ones can be hardly seen.


First, we comment on the curves corresponding to GW experiments.
The shapes of the curves reflect the behavior of the spectrum 
when the values of $\alpha$ or $G\mu$ are varied as described 
in Sec.~\ref{sec:GW}.
Some curves are horizontal for small $\alpha$.
This is because loops cannot emit GWs detectable 
by the corresponding experiments and only the GW background produced 
by kinks on infinite strings contibute.
As already mentioned, for smaller $G\mu$, the effect of particle 
emission is more crucial.
It may be apparently strange that in the curves of BBO and LIGO, 
larger $\alpha$ region is more sensitive to the effect of particle 
emission, since particle emission is more important for smaller loops.
However, for smaller $\alpha$, GWs of frequency to which each detector 
is sensitive were emitted later from larger loops, 
as shown in Fig.~\ref{fig:Omegagwalpha}. 

The excluded region from BBN is separated into two areas.
The triangular region appearing in the right side of 
Fig.~\ref{fig:alphaGmuall} is in the region where loops are long-lived.
In the upper half of it GW emission dominates over particle emission 
(case (3) in subsection \ref{sec:BBN}), while in the lower half the 
reverse holds true (case (4)). 
In the quadrangular region at the upper-left corner, loops are 
short-lived.
Near the right edge GW emission is dominant (case (1)) and
near the lower edge particle emission is dominant (case (2)).

As explained before, the constraint from LSP overproduction is 
similar to that from BBN in that both constrain the visible energy 
injection. 
However, the former is more sensitive to the energy injection 
at the LSP freezeout.
Since the LSP freezeout takes place much earlier than BBN, 
the loop size is smaller and particle productions tend to be dominant 
at that time.
This excludes the large parameter region as shown 
in Fig.~\ref{fig:alphaGmuall}.
On the boundary of the prohibited region, the horizontal part is in the 
region where loops are short-lived and particle emission dominated.
On the other hand, on the downward-sloping part they are long-lived, 
while particle emission still dominates.
Note that in Fig.~\ref{fig:alphaGmuall}, the constraint from LSP 
overproduction is obtained assuming the LSP freezes out in the radiation 
dominated era.
If we use (\ref{T_RH}) as the reheating temperature, the lower boundary
of the prohibited region by LSP overproduction becomes nearly horizontal 
over whole range of $\alpha$, since LSP freeze-out occurs during 
oscillation of flaton and particle emission dominates over GW emission 
for the parameters near the lower boundary 
(it can be seen from (\ref{delrho_RH}) that the energy of LSPs emitted 
at freeze-out depends on only $G\mu$ when loops are particle emission 
dominated).
It is interesting that the tension is much more severely constrained 
than the bound from CMB ($G\mu \lesssim 5\times 10^{-7}$).
This limits models of SSB with a soft SUSY mass term producing 
cosmic strings.
Furthermore, the BBN and DM bounds put considerably severer constraints than the current pulsar timing experiments.

\begin{figure}[t]
\begin{center}
\includegraphics[width=120mm]{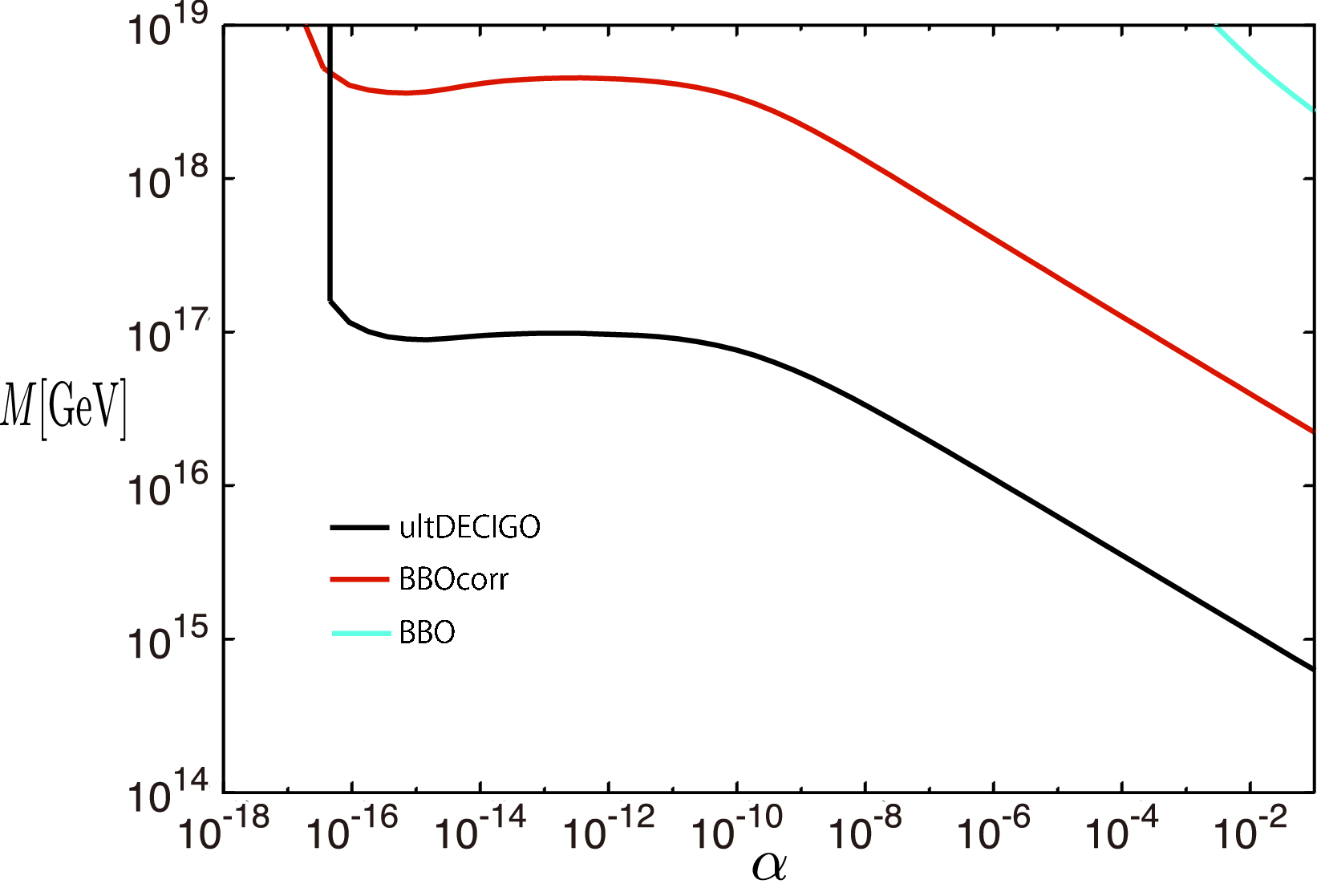}
\caption{
   Parameter regions excluded by cosmological constraints and 
   probed by future experiments on $\alpha$-$M$ plane in the case of 
   $n=2$.
   The meaning of the lines are same as those 
   in Fig.~\ref{fig:alphaGmuall}. 
}
\label{fig:alphaMC}
\end{center}
\end{figure}

\begin{figure}[h]
\begin{center}
\includegraphics[width=120mm]{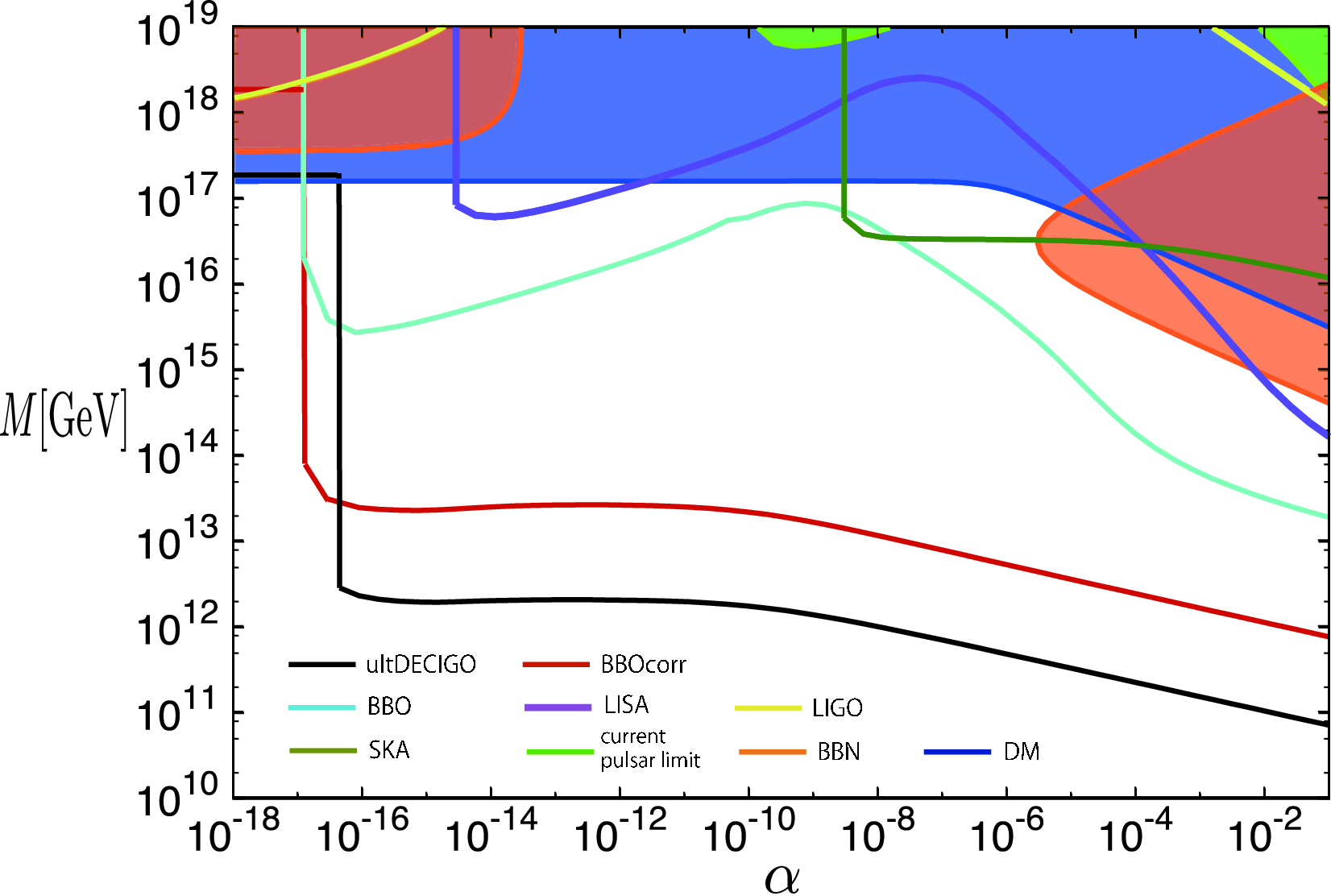}
\caption{
   The same as Fig.~\ref{fig:alphaGmuall} but for $n=3$. 
}
\label{fig:alphaMD}
\end{center}
\end{figure}

The constraints in Fig.~\ref{fig:alphaGmuall} can be translated into 
those on the parameters 
$M$ in Eq.~(\ref{WNR}) and $\alpha$.
The results for $n=2$ or $n=3$ are shown in Fig.~\ref{fig:alphaMC} 
and \ref{fig:alphaMD}, respectively.   
As discussed in Sec.~\ref{sec:model}, it is difficult to raise 
the tension as high as the GUT scale keeping the width to be TeV scale 
for $n=2$.
Therefore, cosmic strings in the model of $n=2$ can be searched 
only by future space-borne interferometers as long as $M$ is 
at most the Planck scale.
On the other hand, for $n=3$, the tension of cosmic strings can be 
as large as the GUT scale, and hence various experiments and 
cosmological bounds set stringent constraints on $M$ and $\alpha$.


\section{Conclusion}   
\label{sec:conc}

In this paper, we have investigated cosmological constraints on 
cosmic strings with TeV-scale width.
We have seen that this type of strings naturally appears 
when the scalar field is the flat direction and it has a soft SUSY 
breaking negative mass-squared term, since the scale of the soft 
mass $m$ is $\mathcal{O}({\rm TeV})$.
Loops arising from such cosmic strings lose the significant part 
of their energies through particle emission from cusps, in addition 
to GW emission.
We have considered effects of the particle production on BBN and DM 
abundance, and the stochastic GW background produced by loops 
including the effect of particle production into the evolution of loops.
We have derived constraints on the tension $\mu$ and the loop size 
$\alpha$ from BBN, DM, and current 
pulsar timing limits and also found the parameter region which can 
be probed by future experiments.
From the result, we have seen that the effects of particle emission 
on the GW spectrum are important for future GW experiments and 
for the BBN and DM bounds for the cosmic strings with TeV-scale width.
Specifically, future space-based GW detectors cover 
very wide parameter ranges of interest.

\appendix

\section*{Appendix}

\section{Energy loss from a cusp} \label{sec:app}

Let us write the string trajectory as $\mathbf{x}(\sigma,t)$ where $\sigma$ is the spacelike parameter on the string.
The most general solution to the Nambu-Goto action is given by~\cite{Vilenkin}
\begin{equation}
	\mathbf{x}(\sigma,t)=\frac{1}{2}
	   \left\{ \mathbf{a}(\sigma-t)+\mathbf{b}(\sigma+t)\right\},
\end{equation}
with constraint $|\mathbf{a'}|=|\mathbf{b'}|=1$.
The cusp is defined as the point where $\mathbf{a'}+\mathbf{b'}=0$.
The coordinate is taken so that $\mathbf{x}=0, \sigma=0$ at the cusp.
Near the cusp, it can be expanded as
\begin{equation}
	\mathbf{x}(\sigma) \simeq \frac{1}{2}\mathbf{x_0''}\sigma^2 
	   + \frac{1}{6}\mathbf{x_0'''}\sigma^3,
\end{equation}
where the subscript 0 means the value evaluated at $\sigma=0$.
Its time derivative is written as
\begin{equation}
	\dot{\mathbf{x}}(\sigma) \simeq 
	\frac{1}{2}(-\mathbf{a_0'}+\mathbf{b_0'})
	+\frac{1}{2}(-\mathbf{a_0''}+\mathbf{b_0''})\sigma
	+ \frac{1}{4}(-\mathbf{a_0'''}+\mathbf{b_0'''})\sigma^2.
\end{equation}
From the condition $|\mathbf{a'}|=|\mathbf{b'}|=1$, we soon obtain
$\mathbf{a_0'} \mathbf{a_0''}=0$, 
$|\mathbf{a_0''}|^2+\mathbf{a_0'}\mathbf{a_0'''}=0$
and similarly for $\mathbf{b_0}$.
Using them, we obtain 
$|\mathbf{\dot x}|^2 \simeq 1-|\mathbf{x_0''}|^2 \sigma^2$.
Thus the the Lorentz factor near the cusp is estimated as
\begin{equation}
	\gamma = \frac{1}{\sqrt{ 1- |\mathbf{\dot x}|^2}} \simeq \frac{1}{|\mathbf{x_0''}|\sigma}.
\end{equation}

Let us estimate the overlap region near the cusp.
Taking into account the Lorentz contraction of the string, its radius 
is given by $\sim w/ \gamma$, where $w$ is the string width.
Notice that the cusp direction is parallel to $\mathbf x_0''$ and 
orthogonal direction, which coincides with the direction of the cusp 
motion, is parallel to $\mathbf{x_0'''}$.
Thus the overlap region is given by 
$-\sigma_c \lesssim \sigma \lesssim  \sigma_c$,
where $\sigma_c = [w/ (\gamma |\mathbf{x_0'''}|)]^{1/3}$.
Substituting $|\mathbf{x_0''}|\sim 1/l$ and 
$|\mathbf{x_0'''}|\sim 1/l^2$, we obtain
$\sigma_c \sim \sqrt{w l}$, where $l$ is the loop radius.
Assuming that the field annihilates efficiently and emit particles 
in the overlap region, the total energy emitted in the form of visible 
particles at the cusp is estimated as
\begin{equation}
	E_{\rm cusp} = \mu \sigma_c \simeq \mu (w l)^{1/2}.
\end{equation}
Since the frequency of cusp formation is $\sim 1/l$, 
the power radiated by the cusp is
$\dot E_{\rm cusp}\simeq \mu (w/l)^{1/2}$~\cite{BlancoPillado:1998bv}.


\section*{Acknowledgment}

This work is supported by Grant-in-Aid for Scientific research from
the Ministry of Education, Science, Sports, and Culture (MEXT), Japan,
No.\ 14102004 (M.K.), No.\ 21111006 (M.K. and K.N.), No.\ 22244030 (K.N.), No.\ 23.10290 (K.M.)
and also by World Premier International Research Center
Initiative (WPI Initiative), MEXT, Japan.


{}

\end{document}